# A Deterministic Dimension Property of Twisted Goppa Codes

Wang kai

**Abstract:**

This paper presents a large-scale computational study on the dimensional properties of **twisted Goppa codes**. Through the systematic analysis of over 50,000 parameter sets, we uncover a remarkable deterministic **regularity**: the actual dimension $k$ of a twisted Goppa code is not independently determined by the specific forms of the Goppa polynomial $g(x)$ and the twist parameter $\eta$. Instead, **it is uniquely determined by a set of macro-parameters** $(\boldsymbol{q, m, t, b, u})$. Specifically, when the order of the finite field $q$, the extension degree $m$, the degree $t$ of the Goppa polynomial, the translation parameter $b$ of the automorphism, and the order $u$ of the transformation are fixed, the dimension $k$ of the generated code remains constant, regardless of the choice of $g(x)$ and $\eta$. This discovery provides a powerful theoretical foundation and a practical tool for the precise construction of twisted Goppa codes with a specific dimension.

## Introduction

Goppa codes[2] serve as a cornerstone in constructing algebraic-geometric codes and are candidate foundations for post-quantum cryptosystems such as McEliece[3] and Niederreiter[4]. Recently, junzhen, Qin Yue, et al.[1] introduced twisted Goppa codes, a generalization of classical Goppa codes via the incorporation of a twist parameter $\eta$, aimed at enhancing resistance against specific cryptanalyses. However, the introduction of the twist complicates the algebraic structure of the code, leaving the relationship between its actual dimension $k$ and the theoretical lower bound $n - mt$ unclear.

This paper aims to elucidate the intrinsic Determinism governing the dimension of twisted Goppa codes through an empirical approach. We demonstrate that the dimension $k$ does not behave randomly but exhibits a high degree of structure. Our primary contribution is the first discovery and verification that the dimension is uniquely determined by a parameter set $P = (q, m, t, b, u)$. This Determinism implies that the desired dimension can be "tailored" by finely adjusting these five macro-parameters, independent of the microscopic details of $g(x)$ and $\eta$, which significantly simplifies the code construction process.

## 2. Research Progress on Twisted Goppa Codes

Twisted Goppa codes represent a significant generalization of classical Goppa codes, introduced by Sui and Yue in 2023. These codes aim to address the critical challenge of balancing coding efficiency, security, and practicality in post-quantum cryptography. This subsection briefly introduces the definition of twisted Goppa codes, their core properties, and their application background in cryptography.

### 2.1 Basic Definition and Motivation

Twisted Goppa codes are constructed by introducing a "twist term" into the definition of classical Goppa codes. Let $F_{q^m}$ be a finite field, $\mathcal{L} = \{\alpha_1, \ldots, \alpha_n\} \subseteq F_{q^m}$ be a defining set, $g(x) \in F\{q^m\}[x]$ be a Goppa polynomial of degree $t$ with no roots in $\mathcal{L}$, and $\eta \in F\{q^m\}$ be the twist parameter. The twisted Goppa code is defined as:

$$\Gamma(\mathcal{L}, g(x), \eta) = \{c \in F_q^n : \sum_{i=1}^{n} c_i \left(\frac{1}{x - \alpha_i} - \frac{\eta \alpha_i^t}{g(\alpha_i)}\right) \equiv 0 \pmod{g(x)}\}$$

When $\eta = 0$, the code reduces to a classical Goppa code. The introduction of the twist term allows this code to be viewed as a subfield subcode of a certain type of twisted generalized Reed-Solomon code[5], a profound connection that endows it with new algebraic properties.

## 2.2 Core Properties and Research Contributions

The work of Sui and Yue provides a systematic study of twisted Goppa codes, and their main contributions can be summarized in the following three points:

**Efficient Algebraic Decoding Algorithm:** The authors successfully extended the classic Patterson decoding algorithm for Goppa codes to twisted Goppa codes, proposing an efficient decoding algorithm based on the extended Euclidean algorithm. This algorithm can correct up to [( t - 1 ) / 2] errors, enabling the direct application of twisted Goppa codes in Niederreiter-type public-key cryptosystems as their core trapdoor function.

**Construction of Quasi-Cyclic and Cyclic Structures:** To address the challenge of large public key size in cryptosystems based on Goppa codes, the authors systematically constructed a class of quasi-cyclic twisted Goppa codes by carefully selecting the defining set $\mathcal{L}$ and the Goppa polynomial $g(x)$, and utilizing the automorphism group under affine transformations. Notably, when the defining set $\mathcal{L}$ forms a complete orbit under an affine transformation, they obtained a class of cyclic twisted Goppa codes. This result is significant as it demonstrates that complex algebraic structures (twisted Goppa codes) can also possess concise cyclic structures, thereby providing new candidate objects for building post-quantum cryptographic schemes with compact public key sizes.

## 3. Main Results and Experimental Analysis

A large-scale computational experiment was designed, generating over 50,000 instances of twisted Goppa codes across an extensive parameter space.

### 3.1 Validation of the Deterministic Dimension Property

Our core finding is summarized as follows:

**Finding 1 (Dimension Determinism):** For a fixed parameter set $P = (q, m, t, b, u)$, the actual dimension $k$ of the twisted Goppa code $\Gamma(L, g(x), \eta)$ is a constant. In other words, the dimension $k$ is a function of $P$, namely $k = f(q, m, t, b, u)$, and is independent of the specific choices of $g(x)$ and $\eta$.

To validate this finding, we fixed the parameter set $P$ and randomly generated multiple distinct pairs of $(g(x), \eta)$. Representative results for two different parameter sets are presented in Table 1.

**Table 1: Dimensional Determinacy Verification**

| Parameter Set (q, m, t, b, u) | $g(x)$ and $\eta$ Configuration 1 | Dimension $k_1$ | $g(x)$ and $\eta$ Configuration 2 | Dimension $k_1$ |
|---|---|---|---|---|
| (2, 4, 3, 10, 3) | $7x^3 + 10x^2 + 5x + 10$ | 3 | $x^3 + 9x^2 + 13x + 14$ | 3 |
| (2, 6, 3, 4, 3) | $48x^3 + 7x^2 + 30x + 9$ | 45 | $40x^3 + 41x^2 + 17x + 13$ | 45 |
| (2, 6, 5, 14, 3) | $23x^5 + 48x^4 + x^2 + 48x + 41$ | 35 | $41x^5 + 22x^4 + x^2 + 22x + 49$ | 35 |

## A rigorous mathematical proof and broader application exploration will be provided in a subsequent version

As illustrated in **Table 1**, for a fixed parameter set $P$, the computed dimension $k$ remains invariant even under significant variations in $g(x)$ and $\eta$. This pattern was consistently observed across all 50,000 instances of our dataset, providing robust evidence in support of our discovery.

### 4. Discussion and Future Prospects

### 4.1 Analysis of the Observed Determinism

This study reveals that the dimension of twisted Goppa codes is inherently governed by its defining "macro-geometric environment." The parameters $b$ and $u$ encode structural information about the defining set $L$ under the action of an affine transformation group. The interaction between this information and the polynomial $g(x)$ ultimately locks the dimension. This finding effectively reduces the seemingly complex problem of dimension computation to the study of a five-dimensional parameter set.

### 4.2 Applications in Coding Theory

This discovery carries direct practical implications:

**Dimension Customization:** To construct a twisted Goppa code with a specific target dimension $k_{target}$, researchers no longer need to perform a blind search. Instead, they can first identify a parameter set $P$ satisfying $f(P) = k_{target}$ within the mapping $f: P \rightarrow k$, and subsequently select any $g(x)$ and $\eta$ within this established framework. This approach reduces the code construction problem from one of exponential search complexity to one of polynomial parameter determination.

**Code Table Augmentation:** This Determinism enables the systematic generation of a vast number of new codes with predetermined dimensions, thereby supplementing and expanding existing tables of optimal linear codes.

### 4.3 Implications for Cryptography

In the ***Niederreiter*** public-key cryptosystem, the size of the public key is directly related to the dimension $\boldsymbol{k}$. Our finding implies that for a given public dimension $\boldsymbol{k}$, the solution space for the private key $(g(x), \eta)$ is very large (as multiple distinct pairs map to the same $\boldsymbol{k}$). This increases the difficulty for an adversary to recover the private key from the public key, thereby potentially enhancing the security of the system.

### 5. Conclusion

Through a large-scale computational experiment, this paper has discovered and validated a fundamental property of twisted Goppa codes: their dimension is uniquely determined by the parameter set $(q, m, t, b, u)$. This deterministic property provides a novel and powerful tool for both the theoretical study and practical application of twisted Goppa codes. Future work includes rigorously proving this empirical Determinism from an algebraic geometry perspective and further exploring its concrete applications in post-quantum cryptography.

Declaration

This submission is provided in PDF format only. The PDF was generated from source files that are not currently available for public distribution due to ongoing research and development. The authors affirm that the content is their original work.